\documentclass{elsart}
\usepackage{epsfig}
\usepackage{amsmath}
\usepackage{url}
\begin{document}
\begin{frontmatter}
\title{Measuring Type Ia Supernova Distances and Redshifts From Their
Multi-band 
Light Curves}
\author[add1]{Alex G. Kim}
\author[add2]{Ramon Miquel}
\address[add1]{Physics Division, Lawrence Berkeley National Laboratory,
Berkeley, CA\ 94720, USA}
\address[add2] {Instituci\'o Catalana de Recerca i Estudis Avan\c{c}ats (ICREA) \\
Institut de F\'{\i}sica d'Altes Energies (IFAE) \\
Edifici Cn, Campus UAB, E-08193 Bellaterra (Barcelona), Spain}

\begin{abstract}
The distance and redshift of a type Ia supernova can be determined
simultaneously through its multi-band light curves.  This fact may be
used for imaging surveys that discover and obtain photometry for large
numbers of supernovae; so many that it would be difficult to obtain a
spectroscopic redshift for each.  Using available supernova-analysis
tools we find that there are several conditions in which a viable
distance--redshift can be determined.  Uncertainties in the effective
distance at $z \sim 0.3$ are dominated by redshift uncertainties
coupled with the steepness of the Hubble law.  By $z
\sim 0.5$ the Hubble law flattens out and distance-modulus
uncertainties dominate. Observations that give $S/N=50$ at peak
brightness and a four-day observer cadence in each of $griz$-bands are
necessary to match the intrinsic supernova magnitude dispersion out to
$z=1.0$.  Lower $S/N$ can be tolerated with the addition of redshift
priors (e.g.\ from a host-galaxy photometric redshift), observations
in an additional redder band, or by focusing on supernova redshifts
that have particular leverage for this measurement.  More stringent
$S/N$ requirements are anticipated as improved systematics control
over intrinsic color, metallicity, and dust is attempted to be drawn
from light curves.
\end{abstract}

\end{frontmatter}

\section{Introduction}
Proposed wide-field imaging surveys will be able to discover and build
light curves of thousands to hundreds of thousands of high-redshift
type Ia supernovae (SNe Ia).  SNe Ia are established as excellent
distance indicators having been used for both for the measurement of
the Hubble Constant \cite{Freedman:2001} and for the discovery of the
accelerated expansion of the Universe
\cite{Riess_acc_98,42SNe_98}.  There is thus interest
in exploring how supernovae in new surveys can be used to improve the
measurement of the expansion history of the Universe and provide
further insight into the physical cause of its acceleration.

Although planned facilities and surveys provide straightforward
harvesting of light curves for large numbers of supernovae, the
corresponding spectroscopic observations used for redshift
determination, supernova typing, and diagnostics are expensive. It is
unclear whether there will be available spectroscopic resources
commensurate to the production of the imaging.  In addition, those
surveys that target spectroscopy for a specific redshift range will
still accumulate supernova light curves at other redshifts
\cite{Alderingetal:2007}.  There is therefore interest in the
possibility of performing supernova cosmology analysis with
photometric data only.  In this scenario, light curves not only fill
their traditional role in measuring distances but are also responsible
for supernova typing and redshift determination.

In this paper, we explore the feasibility of using survey photometry
to simultaneously estimate distance and redshift using fits to a
light-curve template.  We do not incorporate typing and assume that
the supernova is already known to be type Ia.  We use an effective
distance-modulus uncertainty as the metric of interest. Barris and
Tonry \cite{BarrisAndTonry} describe an alternative approach to the
same problem using Bayesian statistics and marginalizing over
redshift, obtaining distance precisions close to the intrinsic
corrected SN Ia magnitude dispersion.  Other papers have considered
other combinations of information derivable from photometric data
only: Johnson and Crotts \cite{JohnsonAndCrotts} and Kuznetsova and
Connolly
\cite{Kuznetsova:2007} for photometric supernova typing; Sullivan et
al.\ \cite{Sullivan:2006} for simultaneous typing and redshift
determination with early light curves; and Wang \cite{Wang:2007} for
supernova photometric redshifts.  The effect of supernova redshift
uncertainties on the determination of dark-energy parameters is
examined in Huterer et al.\ \cite{Huterer&Kim:2004}.

This paper is organized as follows: \S\ref{approach:sec} describes our
approach towards estimating distance modulus and redshift
uncertainties for photometric survey data through light curve fitting.
In \S\ref{spec:sec} we specify the properties of the six surveys that
we consider: all share a four-day cadence but have differing depths
that correspond to obtaining a signal-to-noise (S/N) of either 25 or
50 for a fiducial $z=0.325$, 0.731, or 1.11 supernova at peak
brightness in each band. Interesting features of our results are
discussed in
\S\ref{discussion:sec} and conclusions are presented in
\S\ref{conclusions:sec}.

\section{Approach}
\label{approach:sec}
Our objective is to determine how well the distance and redshift of a
single supernova can be determined from a set of photometric data. We
use a parameterized description of the time-evolving spectral energy
distributions (SEDs) for SNe Ia incident to the observer.  An
individual supernova is characterized by the date of explosion and
redshift in addition to the parameters of the SED model. The survey is
described by the observing cadence, photometric noise, and bands of
observation. Model-parameter uncertainties from light-curve fitting
are estimated using the Fisher information matrix.  These parameters
are in turn propagated into a covariance matrix for the distance
modulus and redshift.  The covariance matrix is distilled into an
effective distance modulus uncertainty to aid in the interpretation of
the results.  The Fisher analysis provides a firm lower limit on the
errors one can obtain: this limit is an excellent estimate of errors
when the data uncertainties are small and Gaussian distributed.

SALT2 \cite{Salt2} provides an empirical model for the time evolving
SED of SNe Ia.  The model is constructed from a training set of
spectroscopic and photometric measurements of both low- and
high-redshift supernovae.  SALT2 models the SED as a function of phase
$p$ and wavelength $\lambda$ by
\begin{equation}
N_{\ln{\lambda}}(p,\lambda;x_0,x_1,c) =  x_0 \times \left[M_0(p,\lambda)+x_1M_1(p,\lambda)\right]  \times e^{c CL(\lambda)}
\label{saltmodel:eqn}
\end{equation}
where the model parameters $x_0$, $x_1$, and $c$ correspond
(approximately) to the peak luminosity, light-curve shape, and
observed color. The color parameterization simultaneously accounts for
intrinsic supernova color variation and foreground dust extinction.
The functions $M_0$, $M_1$, and $CL$ are constructed according to the training
set.  The model covers supernova phases from $\left[-20,+50\right]$ days
and the wavelength range $\left[2000,9200\right]$ \AA.  We work with
photon rather than energy fluxes since they, combined with
transmission functions, describe counter detectors and observer
magnitude systems
\cite{Nugent:2002}.  The use of
densities in $\ln{\lambda}$ rather than $\lambda$ simplifies
the description of redshifted spectra.

The residual dispersion of the training set data from the model gives
a quantitative estimate of how well SALT2 represents supernovae.
SALT2 treats the dispersion of the template light curves of an
average $x_1$=0,
$c=0$ supernova as
\begin{equation}
\sigma_{disp}(N(p, \lambda)) = N(p,\lambda)  \sqrt{ V0(p,\lambda)} DS(p,\lambda) 
\label{residual:eqn}
\end{equation}
where $\lambda$ is the effective wavelength of the band in the
SN-frame and the functions for variance, $V0$, and dispersion scaling,
$DS$, are provided with the software
distribution\footnote{\url{http://supernovae.in2p3.fr/~guy/salt/index.html}}.
Correlations between the residuals within a light curve, though
non-zero, are not quantified.  Dispersion in colors are given
\cite{Salt2} by a wavelength-dependent, phase-independent magnitude
dispersion
\begin{equation}
\sigma_{col}(\lambda)  = 
\begin{cases}
 0.022 \left(\frac{\lambda-\lambda_B}{\lambda_U-\lambda_B}\right)^3 &
 \text{if $\lambda < \lambda_B$},\\ 0.018
 \left(\frac{\lambda-\lambda_V}{\lambda_R-\lambda_V}\right)^2 &
 \text{if $\lambda > \lambda_V$}, \\ 0 & \text{if $\lambda_B \le
 \lambda \le \lambda_V$}.
\end{cases}
\label{residualcolor:eqn}
\end{equation}

The survey data are described by the band of observation and
photometric uncertainty. Each supernova is observed with an
undisrupted cadence in all bands.  Observations are tuned to provide a
fixed signal-to-noise at peak brightness in each band for
an average $x_1=0$, $c=0$ supernova, and assume a
sky dominated background for calculating the noise off-peak.
This
noise model approximately describes data from a ground-based rolling
supernova search (e.g. Astier et al.\ \cite{Astier:2006}).

The supernova light curves are modeled using the SALT2 SED model with
the addition of parameters for the date of explosion $t_0$ and
redshift $z$
\begin{equation}
f(t,X;\mathbf{p}) = \int T_X(\lambda) N_{\ln{\lambda}}\left(\frac{t}{1+z}-t_0,\frac{\lambda}{1+z};x_0,x_1,c\right) d\ln{\lambda}
\label{datamodel:eqn}
\end{equation}
where $T_X$ is the transmission of band $X$. The
complete parameter set for the light-curve model is
$\mathbf{p}=\{t_0,z,x_0,x_1,c\}$.

Parameter uncertainties from light-curve fitting are determined 
through the Fisher information matrix $\mathbf{F}$
\cite{teg_fisher_97}.  Every data point in all useful bands are
included in the calculation.  Note that for the surveys we consider,
photometric noise is independent of the light-curve parameters.  Any
independent redshift prior uncertainties are added to the $F_{zz}$
element.  The covariance matrix of the parameters is the inverse of
the Fisher matrix $\mathbf{F}^{-1}$.

In SALT2, light-curve parameters are used to calculate the distance
modulus through the relation
\begin{equation}
\mu = -2.5\log(x_0)+\alpha_x \times x_1-\beta \times c + \mbox{const}
\end{equation}
with $\alpha=0.13$ and $\beta=1.77$.  The covariance matrix $\mathbf{U}$
of the
parameters  $\boldsymbol{\eta} =\{\mu,
z\}$ is given by
\begin{equation}
U_{ij} = \sum_{k,l} \frac{\partial \eta_i}{\partial p_k}
 \frac{\partial \eta_j}{\partial p_l}(\mathbf{F}^{-1})_{kl}.
\end{equation}

Cosmological parameter fitting can proceed with the likelihood
function dependent on the $\mu-z$ covariance matrix for each
supernova.  However, to avoid analysis of this non-Gaussian likelihood
and provide intuitive insight, we consider the limiting case where
redshift uncertainties are small. Taylor expanding the theoretical
prediction of the distance modulus $\mu_\gamma$ around the observed
redshift, the probability distribution function to first order
is Gaussian with
a $\chi^2$ for
the one parameter
$\mu$ with an effective variance
\begin{equation}
(\sigma_{\mu,eff})^2=U_{\mu\mu}
+\left(\frac{d\mu_\gamma}{dz}\right)^2 U_{zz} + 2 \frac{d\mu_\gamma}{dz} U_{\mu z}.
\label{effmu:eqn}
\end{equation}
(See \S\ref{sec:app} for a sketch of the derivation.)  In this
fashion, uncertainties in $z$ and their correlation with $\mu$ are
incorporated into an effective uncertainty in distance modulus. We use
the distance modulus for a photon-counting magnitude system that is
related to the standard distance modulus by
$\mu_\gamma=\mu-2.5\log{\left(1+z\right)}$.  As $d\mu_\gamma(z) /dz
>0$ for most all cosmologies of interest, positive correlations in
$\mu$ and $z$ increase the effective distance modulus uncertainty.

\section{SN Model and Survey Specification and Results}
\label{spec:sec}
We adopt a baseline description of SALT2 and the supernova survey.
The covariance matrix of the SALT2 templates between points $i$ and
$j$ (with phases $p_i$ and $p_j$ observed in bands $X_i$ and $X_j$) is
based on the residuals given by Equations~\ref{residual:eqn} and
\ref{residualcolor:eqn}:
\begin{equation}
V_{ij}=
\begin{cases}
\sigma_{i,disp} \sigma_{j,disp}\text{sinc}\left(\pi\frac{p_j-p_i}{5 \text{~days}}\right)+\sigma_{i,col}\sigma_{j,col}
& \text{if $X_i=X_j$},\\
0 & \text{if $X_i \ne X_j$}.
\end{cases}
\label{residualcorrelation:eqn}
\end{equation}
We have modeled the day-to-day correlations in the residuals with the
$\text{sinc}(x)=\sin{x}/x$ function with a time scale of 5 days; this
keeps the intrinsic light curves smooth and provides both positive and
negative correlations.  Residuals in different bands are uncorrelated.
Operationally, this covariance matrix of the model dispersion is added
to that of the photometric data.

We consider only data at supernova phases covered by the SALT2 model;
observations around peak brightness provide most of the leverage in
light-curve fits with our sky-limited background.

Observations are modeled as being made in the Megacam $griz$-bands of
the Canada-France-Hawaii Telescope with the transmissions provided
with the SALT2 software distribution.

As SALT2 is not well trained at wavelengths smaller than the $U$-band
and as SNe Ia have little emission at these low wavelengths, we do not
use observations in this regime.  This naturally defines three
specific redshifts to consider, $z=\{0.325, 0.731, 1.118\}$, where the
effective wavelengths of the $gri$-bands in turn match that of the
SN-frame $U$.  The SN-frame wavelengths that correspond to each
observer band are given in Table~\ref{effz:tab}.  The number of data
points for each supernova is $\frac{70
\text{~days}}{\text{cadence}}(1+z)N_{bands}$.
At $z=0.325$ there
are a total 92 data points in $griz$-bands used in the analysis.  At
$z=0.731$ there are 90 points in $riz$-bands and at $z=1.118$ 74
points in $iz$-bands.   When the $z$-band
observes the SN $U$, there is no second band for a color measurement
and the fit is not constrained.

\begin{table}
\caption{Supernova-frame effective wavelengths in microns
for observer $griz$-bands.  The three redshifts in the table
correspond, respectively, to matches of observer $gri$ to
the restframe $U$.\label{effz:tab}}
\begin{tabular}{|c|cccc|}
\hline
redshift&$g$ & $r$ & $i$ & $z$\\ \hline
0.325 & 0.36 & 0.47& 0.58& 0.69\\
0.731 & 0.28 & 0.36& 0.44& 0.53\\
1.118 & 0.23 & 0.29& 0.36& 0.43\\ \hline
\end{tabular}
\end{table}

\begin{figure}
\begin{center}
\epsfig{file=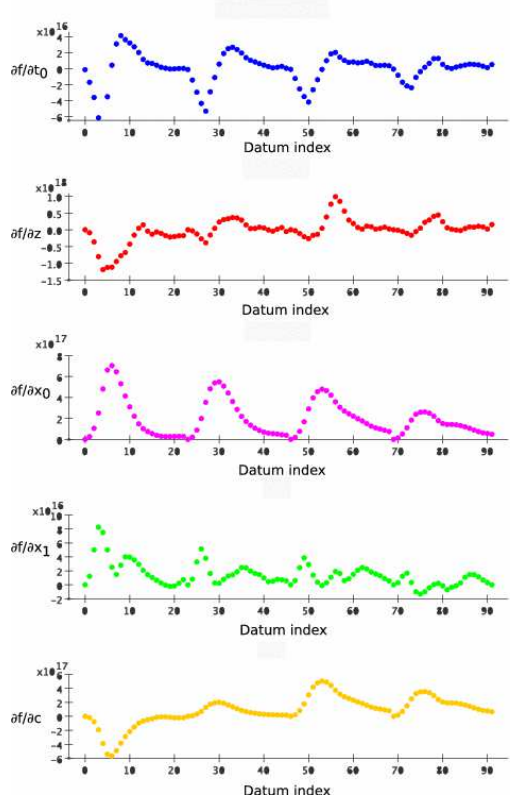,scale=0.4}
\caption{$\partial f/\partial \mathbf{p}$ for the photometric data
of a supernova at $z=0.325$. The indices of the photometric data
points are ordered by the band and by the epoch of observation within
each band.  There are 92 photometric points taken in each band, so the
$n$'th observation in the $\alpha$'th band has index $92\alpha+n$ where
$\alpha=0$ for the $g$, $\alpha=1$ for $r$, $\alpha=2$ for $i$, and
$\alpha=3$ for the $z$ bands.  The units and normalization of $f$ are
set by the SALT2 $x_0=1$ template in MKS units via
Equations~\ref{saltmodel:eqn} and \ref{datamodel:eqn}.  The location
of each point with respect to the light curve is identifiable through
the middle plot $\partial f/ \partial x_0$.
\label{partial0.325:fig}}
\end{center}
\end{figure}

\begin{figure}
\begin{center}
\epsfig{file=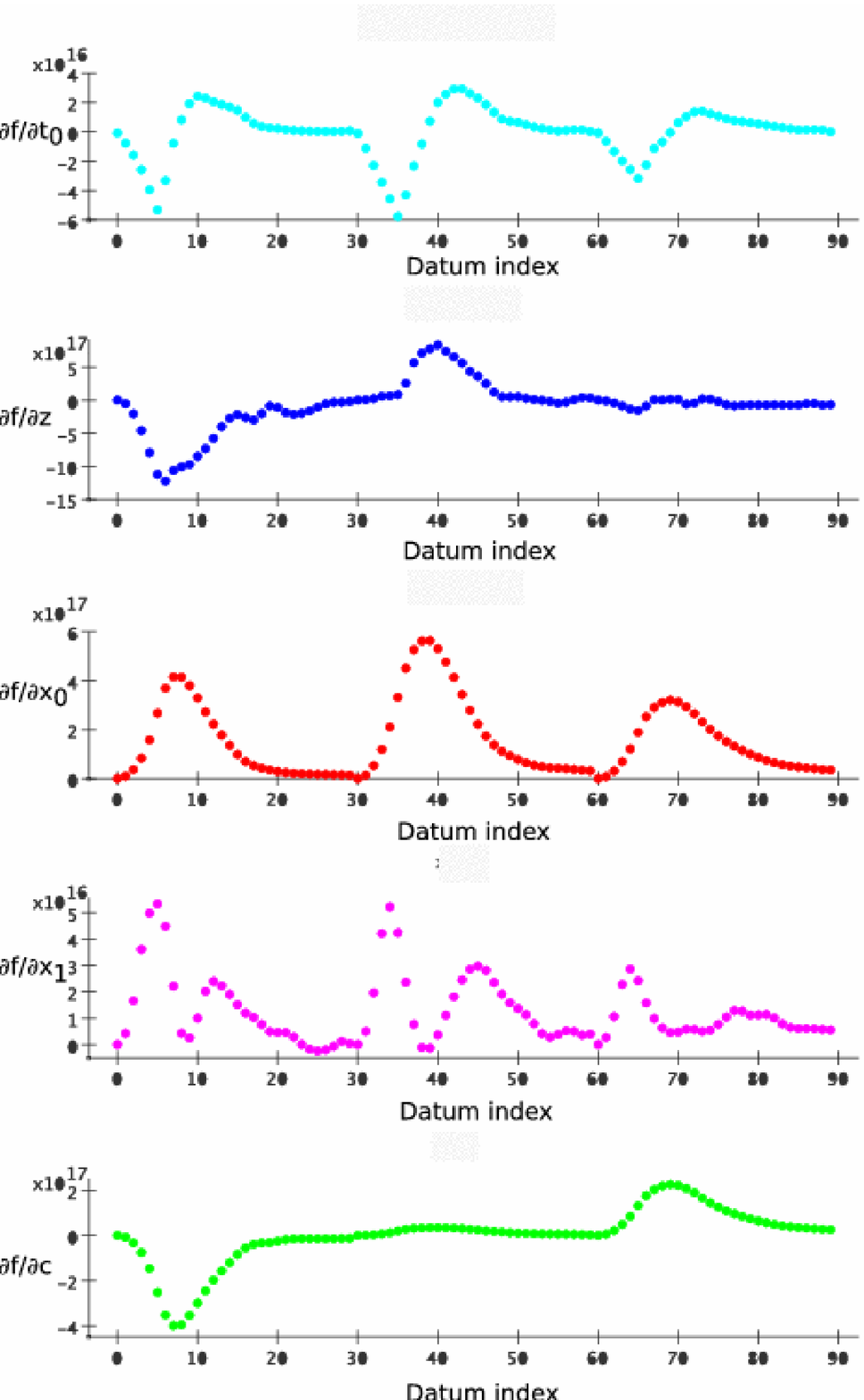,scale=0.4}
\caption{The same as in Figure~\ref{partial0.325:fig} for the 90
data points of a supernova at $z=0.731$. The data correspond, from
left to right, to observations in
$riz$-bands.\label{partial0.731:fig}}
\end{center}
\end{figure}

\begin{figure}
\begin{center}
\epsfig{file=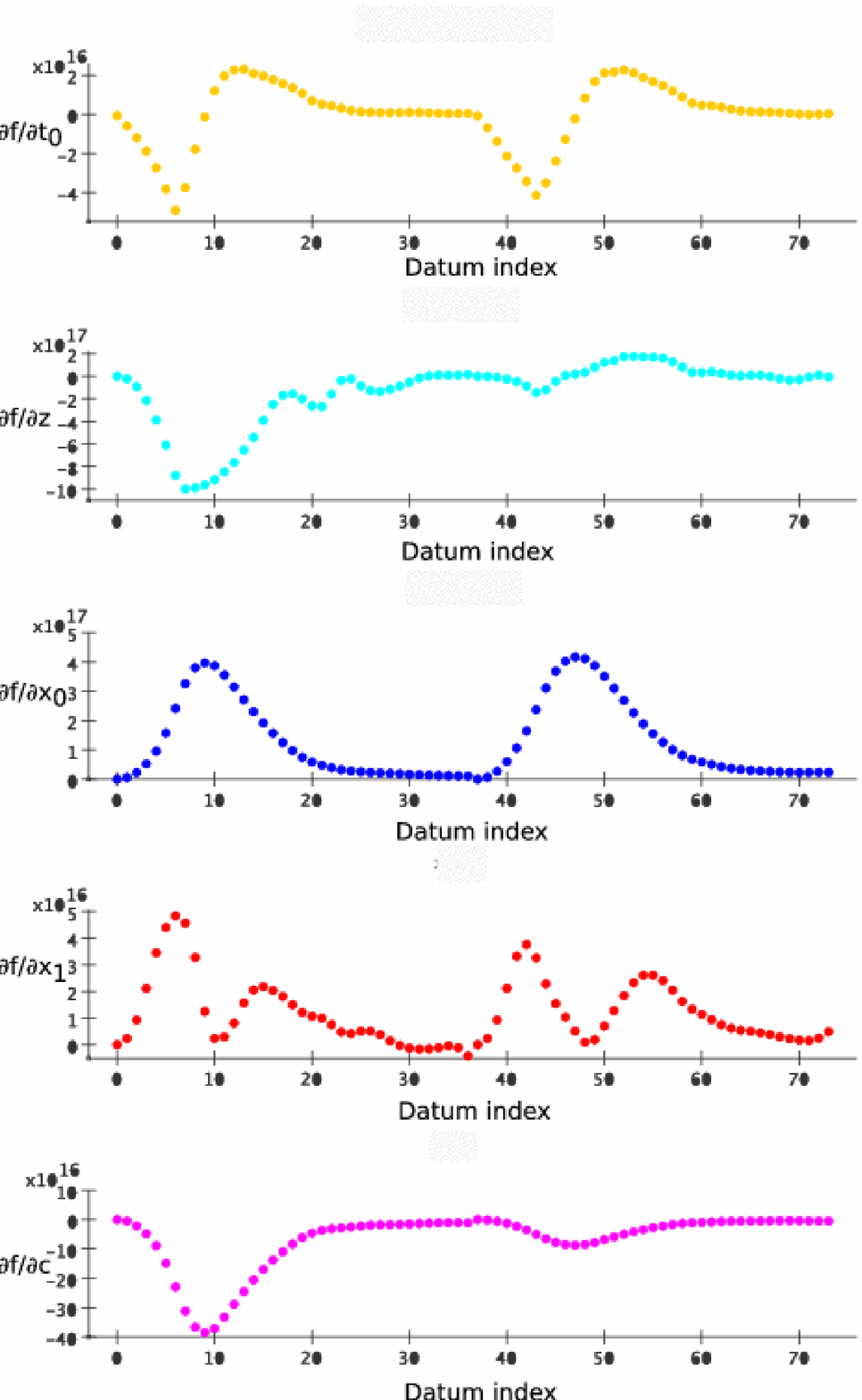,scale=0.4}
\caption{The same as in Figure~\ref{partial0.325:fig} for the
74 data points of a supernova at $z=1.118$. The data correspond, from
left to right, to observations in $iz$-bands.\label{partial1.118:fig}}
\end{center}
\end{figure}

The partial derivatives of the data model of
Equation~\ref{datamodel:eqn} with respect to its parameters are shown
for the three redshifts in Figures~\ref{partial0.325:fig},
\ref{partial0.731:fig}, and \ref{partial1.118:fig}.

We consider surveys designed to provide signal-to-noise of either 25
or 50 for a single visit at the peak brightness of each band for an
average $x_1=0$, $c=0$ supernovae at these specific redshifts. Each
band is observed with a four-day observing cadence.  To simulate the
range of possible spectroscopic or host-galaxy photometric redshifts,
we include redshift priors with precisions of $10^{-3}$, 0.01, as well
as the case of no prior; we expect that supernova photometric
redshifts will easily exclude the extreme non-Gaussian tails from
galactic photometric redshifts.  We use a flat $\Lambda$ universe with
$\Omega_M=0.3$ to get $d\mu_\gamma/dz$ (used in
Equation~\ref{effmu:eqn}) of 6.9, 3.1, and 1.9 for the three
redshifts. The resulting $\mu$--$z$ covariance matrices and effective
$\mu$ uncertainties are given in Table~\ref{sn25:tab} for $S/N=25$ and
Table~\ref{sn50:tab} for $S/N=50$.  Although the results are grouped
according to the $S/N$ at peak brightness, the surveys that generate
these data are different for each redshift: the survey that
corresponds to the high-redshift supernova is deep and obtains higher
$S/N$ for supernovae at lower redshift.

\begin{table}
\caption{$\mu$--$z$ covariance and effective $\mu$ uncertainties: Peak $S/N=25$,
four-day cadence\label{sn25:tab}}
\begin{tiny}
\begin{tabular}{|c|cc|cc|cc|}
\hline
\multicolumn{1}{|c|}{} &\multicolumn{2}{|c|}{$\sigma_{prior}(z)=\text{none}$}  & \multicolumn{2}{|c|}{$\sigma_{prior}(z)=0.01$}    &\multicolumn{2}{|c|} {$\sigma_{prior}(z)=10^{-3}$}\\
\cline{2-7}
\multicolumn{1}{|c|}{$z_{SN}$} &\multicolumn{1}{|c}{$\mathbf{U}$} & \multicolumn{1}{c|}{$\sigma_{\mu,eff}$} &\multicolumn{1}{|c}{$\mathbf{U}$}
& \multicolumn{1}{c|}{$\sigma_{\mu,eff}$} &\multicolumn{1}{|c}{$\mathbf{U}$}
&  \multicolumn{1}{c|}{$\sigma_{\mu,eff}$}\\ \hline
0.32& $ \begin{pmatrix}0.0021& 3.8E-4\\3.8E-4& 2.5E-4\end{pmatrix}$ & $0.14$
& $ \begin{pmatrix}0.0017& 1.1E-4\\1.1E-4& 7.1E-5\end{pmatrix}$ & $0.082$
& $ \begin{pmatrix}0.0015& 1.6E-6\\1.6E-6& 1.0E-6\end{pmatrix}$ & $0.040$
\\
0.731& $ \begin{pmatrix}0.0023& 2.0E-4\\2.0E-4& 8.5E-5\end{pmatrix}$ & $0.065$
& $ \begin{pmatrix}0.0021& 1.1E-4\\1.1E-4& 4.6E-5\end{pmatrix}$ & $0.056$
& $ \begin{pmatrix}0.0018& 2.3E-6\\2.3E-6& 9.9E-7\end{pmatrix}$ & $0.043$
\\
1.118& $ \begin{pmatrix}0.10& 0.012\\0.012& 0.0016\end{pmatrix}$ & $0.39$
& $ \begin{pmatrix}0.017& 7.0E-4\\7.0E-4& 9.4E-5\end{pmatrix}$ & $0.14$
& $ \begin{pmatrix}0.012& 7.5E-6\\7.5E-6& 1.0E-6\end{pmatrix}$ & $0.11$
\\ \hline
\end{tabular}
\end{tiny}
\end{table}

\begin{table}
\caption{$\mu$--$z$ covariance and effective $\mu$ uncertainties: Peak $S/N=50$,
four-day cadence\label{sn50:tab}}
\begin{tiny}
\begin{tabular}{|c|cc|cc|cc|}
\hline
\multicolumn{1}{|c|}{} &\multicolumn{2}{|c|}{$\sigma_{prior}(z)=\text{none}$}   & \multicolumn{2}{|c|}{$\sigma_{prior}(z)=0.01$}    &\multicolumn{2}{|c|} {$\sigma_{prior}(z)=10^{-3}$}\\
\cline{2-7}
\multicolumn{1}{|c|}{$z_{SN}$} &\multicolumn{1}{|c}{$\mathbf{U}$} & \multicolumn{1}{c|}{$\sigma_{\mu,eff}$} &\multicolumn{1}{|c}{$\mathbf{U}$}
& \multicolumn{1}{c|}{$\sigma_{\mu,eff}$} &\multicolumn{1}{|c}{$\mathbf{U}$}
&  \multicolumn{1}{c|}{$\sigma_{\mu,eff}$} \\ \hline
0.325& $ \begin{pmatrix}0.0011& 1.9E-4\\1.9E-4& 1.2E-4\end{pmatrix}$ & $0.097$
& $ \begin{pmatrix}9.3E-4& 8.9E-5\\8.9E-5& 5.4E-5\end{pmatrix}$ & $0.069$
& $ \begin{pmatrix}7.9E-4& 1.6E-6\\1.6E-6& 9.9E-7\end{pmatrix}$ & $0.029$
\\
0.731& $ \begin{pmatrix}0.0010& 7.3E-5\\7.3E-5& 4.4E-5\end{pmatrix}$ & $0.043$
& $ \begin{pmatrix}0.0010& 5.0E-5\\5.0E-5& 3.0E-5\end{pmatrix}$ & $0.040$
& $ \begin{pmatrix}9.3E-4& 1.6E-6\\1.6E-6& 9.8E-7\end{pmatrix}$ & $0.031$
\\
1.118
& $ \begin{pmatrix}0.044& 0.0048\\0.0048& 6.4E-4\end{pmatrix}$ & $0.25$
& $ \begin{pmatrix}0.013& 6.5E-4\\6.5E-4& 8.6E-5\end{pmatrix}$ & $0.12$
& $ \begin{pmatrix}0.0079& 7.5E-6\\7.5E-6& 1.0E-6\end{pmatrix}$ & $0.089$
\\
\hline
\end{tabular}
\end{tiny}
\end{table}

There is no guarantee that SNe Ia can be better standardized than
currently done by SALT2.  However, the optimist may imagine that with
enough data and refinement, the SALT2 model could eventually perfectly
describe supernovae in at all wavelengths.  Table~\ref{sn25nonoise:tab}
shows the improved results when no model dispersion (i.e.\ no contribution
from Equation~\ref{residualcorrelation:eqn}) is included in the
error budget.

\begin{table}
\caption{$\mu$--$z$ covariance and effective $\mu$ uncertainties: Peak $S/N=25$,
four-day cadence, no model residuals\label{sn25nonoise:tab}} 
\begin{tiny}
\begin{tabular}{|c|cc|cc|cc|}
\hline
\multicolumn{1}{|c|}{} &\multicolumn{2}{|c|}{$\sigma_{prior}(z)=\text{none}$}   & \multicolumn{2}{|c|}{$\sigma_{prior}(z)=0.01$}    &\multicolumn{2}{|c|} {$\sigma_{prior}(z)=10^{-3}$}\\
\cline{2-7}
\multicolumn{1}{|c|}{$z_{SN}$} &\multicolumn{1}{|c}{$\mathbf{U}$} & \multicolumn{1}{c|}{$\sigma_{\mu,eff}$} &\multicolumn{1}{|c}{$\mathbf{U}$}
& \multicolumn{1}{c|}{$\sigma_{\mu,eff}$} &\multicolumn{1}{|c}{$\mathbf{U}$}
&  \multicolumn{1}{c|}{$\sigma_{\mu,eff}$}\\ \hline
0.325 & $ \begin{pmatrix}6.7E-4& 1.4E-4\\1.4E-4& 1.4E-4\end{pmatrix}$ & $0.097$
& $ \begin{pmatrix}5.9E-4& 5.9E-5\\5.9E-5& 5.9E-5\end{pmatrix}$ & $0.065$
& $ \begin{pmatrix}5.3E-4& 9.9E-7\\9.9E-7& 9.9E-7\end{pmatrix}$ & $0.024$
\\
0.731& $ \begin{pmatrix}0.0016& 1.6E-4\\1.6E-4& 5.1E-5\end{pmatrix}$ & $0.055$
& $ \begin{pmatrix}0.0014& 1.1E-4\\1.1E-4& 3.4E-5\end{pmatrix}$ & $0.049$
& $ \begin{pmatrix}0.0011& 3.1E-6\\3.1E-6& 9.8E-7\end{pmatrix}$ & $0.033$
\\
1.11& $ \begin{pmatrix}0.068& 0.0085\\0.0085& 0.0011\end{pmatrix}$ & $0.32$
& $ \begin{pmatrix}0.0100& 6.9E-4\\6.9E-4& 9.2E-5\end{pmatrix}$ & $0.11$
& $ \begin{pmatrix}0.0049& 7.5E-6\\7.5E-6& 1.0E-6\end{pmatrix}$ & $0.070$
\\ \hline
\end{tabular}
\end{tiny}
\end{table}

All results presented in this paper are calculated for the case where
observations are phased to begin on the date of explosion.  In the
case of $S/N=25$ and no redshift priors, results vary by $< 1$\%
depending on the phase of observations with respect to the underlying
light curve.

\section{Discussion}
\label{discussion:sec}
There are several interesting features of our results that we discuss
here in further detail.

Distance modulus and redshift measurements have significant positive
correlation when there are no or weak redshift priors.  The
correlation is dominated by the contribution of $\alpha \times
(\mathbf{F}^{-1})_{x_1 z}$. The positive $(\mathbf{F}^{-1})_{x_1 z}$
is due to the supernova-frame $U$-band, where a strongly
negative $\partial
f/\partial z$ arises from the band coverage over the transition between
the UV-flux dropout to the bright $B$ emission
ubiquitous to SNe Ia.  On the other hand, SALT2 predicts brighter UV
flux for larger $x_1$.  The coverage over Ca H\&K wavelengths is
therefore important for distinguishing supernova redshift and light-curve
shape.

The effective distance modulus uncertainty is more sensitive to
redshift uncertainties at low redshift.  For example, in the case of
no redshift prior, $z=0.325$ and $S/N=25$, the large value of
$\left(d\mu_\gamma/dz\right)^2$ makes it such that $\sigma_{\mu,eff}$
comes almost entirely from the $\left(\frac{d\mu_\gamma}{dz}\right)^2
U_{zz}$ term.  The approximation derived in \S\ref{sec:app} to get
Equation~\ref{effmu:eqn} breaks down as $\sigma_z d\mu_\gamma/dz$ gets
close to or larger than $\sigma_\mu$, which is certainly the case at
low redshifts.  By $z=0.731$ the effective distance uncertainty is
dominated by the $U_{\mu\mu}$ contribution and our approximation
holds.  The $d\mu_\gamma/dz$ term does give the effective distance
modulus uncertainty a dependence on the cosmological parameters.  A
fit could proceed using the full $\mu$--$z$ covariance matrix, or
iteratively with the values of the cosmological parameters updated
using the fit from the preceding iteration.

Without redshift priors, the redshift uncertainties derived from supernova
light curves range from 0.011--0.025
for the case of $S/N=25$.  Significant improvement in
the cosmological utility of each supernova is possible with a
comparable or better redshift prior.  This is illustrated by the case
of $z=1.118$ in Table~\ref{sn25:tab}.  Without priors, the uncertainty
in redshift is $0.025$.  When the redshift is constrained
by a prior with 0.01 uncertainty, the resulting 
uncertainty in $\mu$ drops
precipitously from $0.21$ to $0.11$ mag\footnote{
To give the reader the ability to explore other redshift priors, we
provide the Fisher matrix for the supernova at $z=1.118$:
\begin{equation}
\mathbf{F}= \begin{pmatrix} 
 41.979296 &  135.174711 &   26.890238  & -9.767684 &  -11.22469\\
135.174711 & 9417.796467& -2617.136017& -204.29132 &  2893.102837\\
 26.890238& -2617.136017 & 4501.498662 & 262.329719& -1834.985182\\
 -9.767684 & -204.29132 &   262.329719 &  39.340308 & -110.596575\\
-11.22469 &  2893.102837& -1834.985182& -110.596575&  1244.587705
\end{pmatrix}. \nonumber
\end{equation}
}.
Concern about redshift
uncertainties are superfluous when spectrograph-quality redshifts
$\sigma_{prior}(z)=10^{-3}$ are available.  This shows why the
propagation of redshift uncertainties is of little concern for
experiments with spectroscopic redshifts.

It may seem odd that the covariance matrix $\mathbf{U}$ gives smaller
redshift uncertainties for the $z=0.731$ survey than for the $z=0.325$
survey, particularly since the latter has observations in four bands
rather than three.  By construction, the two surveys give the same
$S/N$ for their respective target redshift and both use comparable
numbers of data points due to the extra time dilation experienced by
the more distant object.  Qualitative comparison of
Figures~\ref{partial0.325:fig} and
\ref{partial0.731:fig} shows that  $\partial f/\partial z$
distinguishes $z=0.325$ from 0.731 supernovae. At the lower redshift,
$\partial f/\partial z$ and $\partial f/\partial c$ have almost
identical shapes whereas at $z=0.731$ they are visibly different in
the third band, allowing for tighter constraints in $z$ and $c$.  The
Fisher matrix calculation gives for $z=0.325$: $\sigma_z=0.016$,
$\sigma_c=0.028$ and $\rho_{zc}=-0.72$ and for $z=0.731$:
$\sigma_z=0.0092$, $\sigma_c=0.021$ and $\rho_{zc}=-0.41$.

Differences in $\partial f/\partial z$ between the two redshifts are
due to differing contributions from
\begin{equation}
-\frac{1}{(1+z)^2}\int
T_X(\lambda) \lambda \frac{\partial N_{\ln{\lambda}}}{\partial \lambda'}\left(p,\lambda'\right) d\ln{\lambda}
\label{redshifting:eqn}
\end{equation}
where $\lambda'=\lambda/(1+z)$. This term reflects changes in the
predicted observed light curves with redshifting of the supernova SED
at fixed phase. (The phase dependent contribution to $\partial
f/\partial z$ strongly resembles $\partial f/\partial t_0$.)  The
different shapes and uneven $\ln{\lambda}$-spacing of the $griz$-bands
give redshift-dependent observed supernova-frame
wavelengths. Equation~\ref{redshifting:eqn} is close to zero when the
band covers the wavelength region with peak emission; at shorter
wavelengths it is negative and at longer wavelengths it is positive.
The pronounced peaks in the $i$-band $\partial f/\partial z$ at
$z=0.325$ and $z=0.731$ are attributable to their SN-frame wavelengths
and the phase-dependent $\frac{\partial N_{\ln{\lambda}}}{\partial
\lambda'}\left(p,\lambda'\right)$.

\begin{figure}
\begin{center}
\epsfig{file=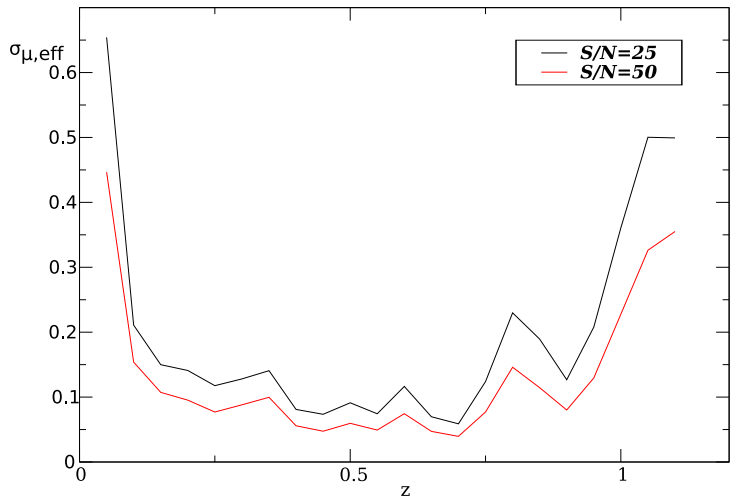,scale=0.4}
\caption{
Calculated $\sigma_{\mu,eff}$ for a series of redshifts
from $z=0.05$ -- 1.1 in steps of 0.05, with $S/N=25$ and 50
photometry and no redshift prior.  The low-frequency
shape of the curve is due to the evolution of $\frac{d\mu_\gamma}{dz}$ at low redshifts, and 
the decreasing amount of data and band coverage at high redshift. The
high-frequency oscillatory behavior is due to differing coverage of
the supernova SED by the observer bands and the resulting change
in sensitivity to the light-curve parameters.
\label{z-dmueff:fig}}
\end{center}
\end{figure}

We calculate $\sigma_{\mu,eff}$ for a series of redshifts
from $z=0.05$ -- 1.1 in steps of 0.05 for the cases
of $S/N=25$ and 50 and no redshift prior, and show the results
in Figure~\ref{z-dmueff:fig}. Looking at the low-frequency behavior,
at low redshift there is a dramatic decrease in  $\sigma_{\mu,eff}$
with increasing redshift as $\frac{d\mu_\gamma}{dz}$ falls.  At redshifts
greater than $z=0.75$, $\sigma_{\mu,eff}$ degrades as the number of
usable data points and band coverage decreases. The
high-frequency oscillatory behavior is due to differing coverage of
the supernova SED by the observer bands and the resulting change
in sensitivity to the light-curve parameters.

Comparison of Tables~\ref{sn25:tab} and \ref{sn25nonoise:tab}
indicates that photometry uncertainty in the $S/N=25$ scenario
dominates over the uncertainty from the SALT2 light-curve templates.
The covariance matrices for the case of no redshift prior differ by a
factor of two between $S/N=25$ and $S/N=50$, rather than the factor of
four expected if there were no SALT2 uncertainties.

Selecting cadences of two, four, and eight days while keeping the peak
signal-to-noise proportional to the square root of the cadence gives
the same results within the second significant digit.  Degradation is
expected at higher cadences as supernova phases with leverage (large
$\partial f/\partial {\mathbf{p}}$) can be missed.

Extending the observer bands to include the $Y$-band can help
significantly for objects at high redshift.  For a $z=1.118$ supernova
with $S/N=25$ at peak, $\sigma_{\mu,eff}$ improves from 0.39 to 0.11
mag.

For the cases in this paper where $\sigma_z>0.01$, we expect that a
full Monte-Carlo-based fit would deviate from the Fisher estimate by
something in the order of 10-20\%; the broad wiggles in the supernova
spectrum produce non-linear effects over this large redshift range.

\section{Conclusions}
\label{conclusions:sec}
In assessing whether a photometric survey produces adequate $\mu$--$z$
uncertainty, we set as a target achieving a $\sigma_{\mu,eff}$
comparable to the intrinsic SN Ia magnitude dispersion, $\sim 0.15$
mag.  At this point, it is more advantageous to spend time observing
more supernovae rather than improving the statistical precision of
each individual object.

If relying on photometric information only, the observer can tune the
depth of the survey to obtain the desired distance
uncertainties.
Measurements that give better than $S/N=50$ at
peak brightness for an average $x_1=0$, $c=0$ supernova in each band
with a four-day observing cadence would be necessary to meet the
target $\sigma_{\mu,eff}$ out to $z=1.0$.
Survey changes that hold fixed $(S/N)/\text{cadence}^2$ have almost no
effect on $\mu$--$z$ uncertainty for the finely sampled light curves
considered in this paper.

Improvements in photometric quality give declining yields when
photometric uncertainties are smaller than those of the SN model.  For
our simulated data, SN model uncertainties begin to dominate between
data qualities corresponding to our $S/N=25$ and $S/N=50$ surveys.
Work on supernova modeling is ongoing and improvements are expected as
data with better wavelength and temporal coverage are included in template
building.

A judicious selection of observer bands and/or targeting of ``magic''
redshifts can give data with stronger leverage in minimizing the
$\mu$--$z$ covariance matrix, as seen in Figure~\ref{z-dmueff:fig}.
These occur when there are strong distinctive light-curve gradients
with respect to the redshift of the spectra, allowing the breaking of
the degeneracy between redshift and other parameters.

Shallower surveys can use the subset of supernovae with redshift prior
uncertainties of $0.01$ or better.  We expect that the patient or
well-equipped observer can obtain spectroscopic redshifts of
supernovae discovered out to $z=0.325$.  Complete spectra for an
unbiased sample of supernova hosts at higher redshifts will be more of
a challenge; for these, photometric redshifts can be used.
Photometric redshift uncertainties of $\sigma_{\Delta z/(1+z)}=0.029$
for a galaxy subsample have been obtained with the CFHT filter set
\cite{Ilbert:2006} and better performance can be obtained with the addition
of a redder filter. A broader and non-Gaussian dispersion is expected,
however, for the diverse population of supernova host galaxies.

In this paper, we do not consider biases in the SALT2 model.  Biases
in the distance and redshift determination leave irreducible
uncertainties that impact cosmological parameter measurements.  The
size of the training set necessary for the SALT2 model construction
and in testing its performance is subject for further study.

As supernova cosmologists concentrate on reducing the systematic
uncertainties in using SNe Ia as distance indicators, focus is being
placed on using light curves to distinguish variation in observed
colors that are intrinsic to supernovae against those due to dust, and
to determine the extinction properties of the host-galaxy dust
itself. Other intrinsic supernova parameters may be encoded in the
light curves.  In addition, we have not considered the impact of
assigning the light-curve fitter the additional responsibility of
determining supernova type.  Constructing a model for such a
generalized fitter is difficult given the heterogeneity of
core-collapse supernovae and the dearth of Ibc light curves, although
Bayesian methods have been applied towards this problem
\cite{BarrisAndTonry,Kuznetsova:2007}.  The formalism presented in
this paper can be applied to future sophisticated light-curve models
with an expanded parameter set.  With photometric information diverted
towards the measurement of additional supernova features, the
uncertainties in $\mu$ and $z$ can only degrade compared to the
3-parameter SALT2 model.

\begin{ack}
We acknowledge helpful discussions with Julien Guy, Eric Linder, Lifan
Wang, and Yun Wang.  Thanks also to the Aspen Center for Physics where
the idea for this paper and much of the work took place.  We thank the
referee for constructive comments. This work has been supported in
part by the Director, Office of Science, Department of Energy under
grant DE-AC02-05CH11231.
\end{ack}

\appendix
\section{Joint Probability Density Function for Distance Modulus and Redshift}
\label{sec:app} 
When attempting to determine the cosmological parameters using the
information contained in the distance modulii and redshifts of a set of
SNe Ia, the joint probability density function (pdf)
$p(\mu^o,z^o | \vec{\theta})$ for the distance modulus ($\mu$) and
redshift ($z$) of each supernova is needed. Here, the superindex $^o$
denotes observed quantities, while $\vec{\theta}$ stands for the
cosmological, and possibly nuisance, parameters.  The pdf is the
probability density for observing a $(\mu^o,z^o)$ pair given
$\vec{\theta}$. It can be computed as
\begin{equation}
\label{eq:1}
p(\mu^o,z^o | \vec{\theta}) = \int\! d\mu^t \int\! dz^t \; p(\mu^o,z^o | \mu^t,z^t)\; p(\mu^t,z^t | \vec{\theta}) \ ,
\end{equation}
where we integrate over all possible true values for $\mu$ and $z$. The first pdf in the right-hand side 
of~(\ref{eq:1}) is just a two-dimensional 
resolution Gaussian relating the observed and the true values of $\mu$ and $z$. The
second pdf can be written as
\begin{eqnarray}
\label{eq:2}
p(\mu^t,z^t | \vec{\theta}) & =     &  p(\mu^t | z^t,\vec{\theta}) \; p(z^t | \vec{\theta})     \nonumber \\ 
                          & =     &  \delta(\mu^t - \overline{\mu}(z^t,\vec{\theta})) \; p(z^t) \nonumber \\
                          &\propto&  \delta(\mu^t - \overline{\mu}(z^t,\vec{\theta})) \ ,
\end{eqnarray}
where $\overline{\mu}(z^t,\vec{\theta})$ is the distance modulus that
results from applying the Hubble relationship to the true redshift,
considering the values of the parameters $\vec{\theta}$. In writing
the last proportionality relationship we have taken into account that
$p(z^t | \vec{\theta})$ does not depend on $\vec{\theta}$ and,
therefore, can be safely neglected in the following. Therefore, we
have
\begin{eqnarray}
\label{eq:3}
p(\mu^o,z^o | \vec{\theta}) &\propto& \int_{-\infty}^{+\infty} d\mu \int_{-\infty}^{+\infty} dz \;
\exp\left[-\frac12\left(\mu^o-\mu,z^o-z\right)^T \mathbf{U}^{-1} \left(\mu^o-\mu,z^o-z\right)\right] 
\delta(\mu - \overline{\mu}(z,\vec{\theta}))\nonumber \\
                         & = & \int_{-\infty}^{+\infty} dz \;
\exp\left[-\frac12\left(\mu^o-\overline{\mu}(z,\vec{\theta}),z^o-z\right)^T
\mathbf{U}^{-1} \left(\mu^o-\overline{\mu}(z,\vec{\theta}),z^o-z\right)\right] \ .
\end{eqnarray}
In the last equation we have dropped the $^t$ superindices for
simplicity, and we have used the delta function to perform the
integral over $\mu$. $\mathbf{U}$ is the $2\times2$ covariance matrix for the
observation of $\mu^o$ and $z^o$.  In order to simplify the
calculations that follow, we will assume for the moment that $\mathbf{U}$ is a
diagonal matrix with elements $\sigma_\mu^2$, $\sigma_z^2$. In this
case, Equation~(\ref{eq:3}) simplifies to:
\begin{equation}
\label{eq:4}
p(\mu^o,z^o | \vec{\theta}) \propto \int_{-\infty}^{+\infty} dz \;
\exp\left[-\frac12\left(\frac{\mu^o-\overline{\mu}(z,\vec{\theta})}{\sigma_\mu}\right)^2\right]
\exp\left[-\frac12\left(\frac{z^o-z}{\sigma_z}\right)^2\right] \ .
\end{equation}
Although the integral extends to all possible values of $z$, it is clear from the second exponential that
only values of $z$ sufficiently close to $z^o$ will actually contribute. Let us, then, expand
$\overline{\mu}(z,\vec{\theta})$ about $z^o$ as
\begin{equation}
\label{eq:5}
\overline{\mu}(z,\vec{\theta}) \simeq \overline{\mu}(z^o,\vec{\theta}) + \left.\frac{d\mu}{dz}\right|_{z^o}(z-z^o) \ .
\end{equation}
Introducing Equation~(\ref{eq:5}) into Equation~(\ref{eq:4}) leads to
\begin{equation}
\label{eq:6}
p(\mu^o,z^o | \vec{\theta}) \sim \int_{-\infty}^{+\infty} dz \;
\exp\left[-\frac{1}{2\sigma_\mu^2}\left(\mu^o-\overline{\mu}(z^o,\vec{\theta})
                                         -\left.\frac{d\mu}{dz}\right|_{z^o}(z-z^o)\right)^2\right]
\exp\left[-\frac12\left(\frac{z^o-z}{\sigma_z}\right)^2\right] \ ,
\end{equation}
which is just a Gaussian integral in $z$. It can be easily computed to give:
\begin{equation}
p(\mu^o,z^o | \vec{\theta}) \sim 
\exp\left[-\frac12\left(\frac{\mu^o-\overline{\mu}(z^o,\vec{\theta})}{\sigma}\right)^2\right] \ .
\end{equation}
So finally the desired pdf is a Gaussian distribution
centered at the expected distance modulus for the measured redshift and with variance
$\sigma^2 = \sigma_\mu^2 + \left(\frac{d\mu}{dz}\sigma_z\right)^2$, where the derivative is to be taken at $z=z^o$.
If the covariance matrix $\mathbf{U}$ between $\mu^o$
and $z^o$ is non-diagonal, the resulting pdf is still a Gaussian with the same mean but with variance
$\sigma^2 = \sigma_\mu^2 + \left(\frac{d\mu}{dz}\sigma_z\right)^2 + 2\rho\frac{d\mu}{dz}\sigma_\mu\sigma_z$, where
$\rho$ is the correlation coefficient between $\mu^o$ and $z^o$.

\end{document}